\documentclass[runningheads]{llncs}

\usepackage[T1]{fontenc}
\usepackage{graphicx}
\usepackage{color}
\usepackage{subcaption}
\usepackage{adjustbox}
\usepackage{multicol}
\usepackage{multirow}
\usepackage{enumitem}
\usepackage{amsmath}

\usepackage{tikz}
\usetikzlibrary{decorations.pathreplacing,calc}

\newcommand{\bigO}[1]{\mathcal{O}(#1)}
\newcommand{\littleO}[1]{{o}(#1)}
\newcommand{\timeComplexity}{$\bigO{\text{ra}(m)+\text{occ})}$}
\newcommand{\spaceComplexity}{$\bigO{\text{er}(T)}$}

\begin{document}

\title{The CDAWG Index and Pattern Matching on Grammar-Compressed Strings}
\titlerunning{The CDAWG Index for SLPs}

\author{Alan M. Cleary\inst{1} \and
Joseph~Winjum\inst{2} \and
Jordan~Dood\inst{3} \and
Shunsuke~Inenaga\inst{4}}

\authorrunning{A. Cleary et al.}
\institute{National Center for Genome Resources, Santa Fe, NM, USA\\
\email{acleary@ncgr.org}\and
Montana State University, Bozeman, MT, USA\\
\email{joseph.winjum@ecat1.montana.edu}\and
Hyalite Technologies LLC, Bozeman, MT, USA\\
\email{hyalitetechnologies@gmail.com}\and
Department of Informatics, Kyushu University, Fukuoka, Japan\\
\email{inenaga.shunsuke.380@m.kyushu-u.ac.jp}}

\maketitle

\begin{abstract}
The compact directed acyclic word graph (CDAWG) is the minimal compact automaton that recognizes all the suffixes of a string.
Classically the CDAWG has been implemented as an index of the string it recognizes, requiring $\littleO{n}$ space for a copy of the string $T$ being indexed, where $n=|T|$.
In this work, we propose using the CDAWG as an index for grammar-compressed strings.
While this enables all analyses supported by the CDAWG on any grammar-compressed string, in this work we specifically consider pattern matching.
Using the CDAWG index, pattern matching can be performed on any grammar-compressed string in \timeComplexity\ time while requiring only \spaceComplexity\ additional space, where $m$ is the length of the pattern, $\text{ra}(m)$ is the grammar random access time, $\text{occ}$ is the number of occurrences of the pattern in $T$, and $\text{er}(T)$ is the number of right-extensions of the maximal repeats in $T$.
Our experiments show that even when using a na\"ive random access algorithm, the CDAWG index achieves state of the art run-time performance for pattern matching on grammar-compressed strings.
Additionally, we find that all of the grammars computed for our experiments are smaller than the number of right-extensions in the string they produce and, thus, their CDAWGs are within the best known $\bigO{\text{er}(T)}$ space asymptotic bound.

\end{abstract}

\section{Introduction}

The compact directed acyclic word graph (CDAWG) is the minimal compact automaton that recognizes all the suffixes of a string.
Classically the CDAWG has been implemented as an index of the string it recognizes, requiring $\littleO{n}$ space for a copy of the string $T$ being indexed, where $n=|T|$.
Grammar-based compression is the method of compressing a string by computing a context-free grammar (CFG) that produces the string (and no others) such that the size of the encoded grammar is smaller than the string.
A CFG that produces a single string is called a straight-line grammar (SLG).

The CDAWG is equivalent to the minified suffix tree for the same string and can be used for a variety of string tasks.
However, the requirement of storing a copy of the original string limits the size of texts the CDAWG can be used on.
Conversely, the SLGs of grammar-compressed strings achieve good compression in practice, especially for highly repetitive strings, but are limited in their ability to perform string analysis tasks.
In this work, we propose utilizing the virtues of both structures by using the CDAWG to index the SLGs of grammar-compressed strings.

This approach will enable all analyses supported by the CDAWG on any grammar-compressed string.
This in itself should be of independent interest as many such analyses have not yet been studied in the context of grammar-compressed strings.
In this work, though, we focus on the problem of pattern matching on grammar-compressed strings.
Using the CDAWG index, pattern matching can be performed on any SLG in \timeComplexity\ time while requiring only \spaceComplexity\ additional space, where $m$ is the length of the pattern, $\text{ra}(m)$ is the grammar random access time, $\text{occ}$ is the number of occurrences of the pattern in $T$, and $\text{er}(T)$ is the number of right-extensions of the maximal repeats in $T$.
Our experiments show that even when using a na\"ive random access algorithm the CDAWG index achieves state of the art run-time performance for pattern matching on grammar-compressed strings.
Additionally, we find that all of the grammars computed for our experiments are smaller than the number of right-extensions in the string they produce and, thus, their CDAWGs are within the best known $\bigO{\text{er}(T)}$ space asymptotic bound.

The rest of the paper is organized as follows:
In Section~\ref{sec:preliminaries} we define syntax and review relevant background information.
In Section~\ref{sec:related} we review related work and in Section~\ref{sec:idea} we discuss how to use the CDAWG to index the SLGs of grammar-compressed strings.
In Section~\ref{sec:algorithms} we discuss  algorithms and data structures. And in Section~\ref{sec:experiments} we discuss our implementation and experimental results. Section~\ref{sec:conclusion} is the conclusion.

\section{Preliminaries}
\label{sec:preliminaries}

In this Section we define syntax and review relevant background information.
Indexes are zero-based.

\subsection{Strings}

Let $T$ be a string over an alphabet $\Sigma$, where $n=|T|$ and $\sigma=|\Sigma|$.
We assume that all strings end with a unique character $\$\notin\Sigma$ that is lexicographically smaller than any character in $\Sigma$.
A \emph{repeat} $\omega$ is a substring of $T$ that occurs at least twice.
A \emph{left-extension} of repeat $\omega$ is a substring $\alpha\omega$ in $T$, where $\alpha\in\Sigma$.
Similarly, a \emph{right-extension} of repeat $\omega$ is a substring $\omega\alpha$ in $T$.
A repeat $\omega$ is \emph{left-maximal} if every left-extension $\alpha\omega$ occurs fewer times in $T$ than $\omega$.
Similarly, a repeat $\omega$ is \emph{right-maximal} if every right-extension $\omega\alpha$ occurs fewer times in $T$ than $\omega$.
A repeat $\omega$ is a \emph{maximal repeat} if it is both left- and right-maximal.
We use $\text{el}(T)$ and $\text{er}(T)$ to denote the number of left- and right-extensions of the maximal repeats in $T$, respectively.

\subsection{CDAWGs}

The compact directed acyclic word graph (CDAWG) is the minimal compact automaton that recognizes all the suffixes of a string~\cite{blumer1987inverted}.
Each edge in a CDAWG has a label representing a non-empty string, and each path from the root node, or \emph{source}, to an internal node has a label that is the concatenation of the labels of the edges in the path.
The longest path to each internal node represents a maximal repeat.
Consequently, there is a one-to-one correspondence between the maximal repeats of a string and the internal nodes of the CDAWG that recognizes the suffixes of the string.
Classically the CDAWG has been implemented as an index of the string it recognizes, meaning start and end positions in the string are used in place of edge labels.
See Figure~\ref{fig:example} for an example CDAWG.

\subsection{Context-Free Grammars}

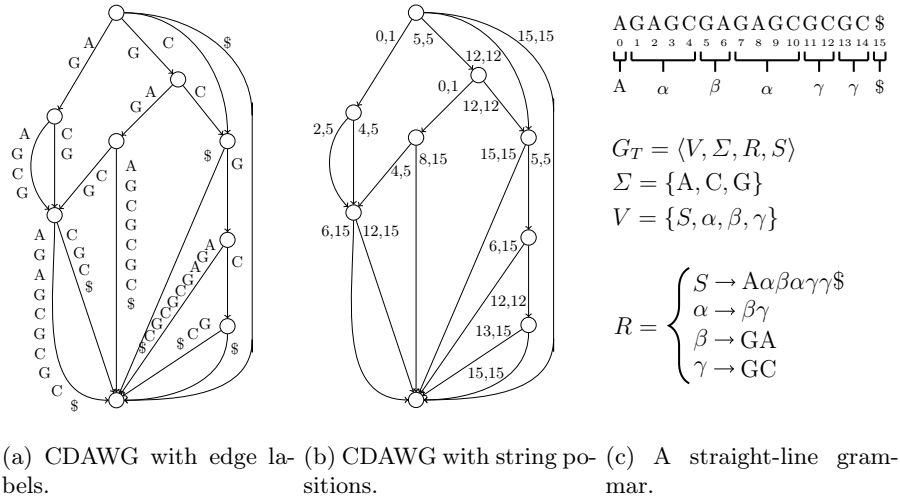
\begin{figure}[b!]
\begin{center}
    \begin{subfigure}[t]{0.32\textwidth}
        \centering
        \begin{adjustbox}{max width=1\textwidth}

\begin{tikzpicture}

    \node[circle, draw] (a) at (0,0.1) {};
    \node[circle, draw] (b) at (1.25,-1.25) {};
    \node[circle, draw] (c) at (0,-2.5) {};
    \node[circle, draw] (d) at (2.25,-2.5) {};
    \node[circle, draw] (e) at (2.25,-4.5) {};
    \node[circle, draw] (f) at (-1.25,-2) {};
    \node[circle, draw] (g) at (2.25,-6.25) {};
    \node[circle, draw] (h) at (-1.25,-4) {};
    \node[circle, draw] (i) at (0,-7.75) {};
    \node[inner sep=0pt] (j) at (2.75,-2) {}; 
    \node[inner sep=0pt] (k) at (2.75,-6.5) {}; 
    \node[inner sep=0pt] (x) at (3.5,-3) {}; 
    
    \node[circle, font=\footnotesize] (10) at (-0.55,-0.5) {A};
    \node[circle, font=\footnotesize] (11) at (-0.85,-0.9) {G};
    \node[circle, font=\footnotesize] (12) at (0.35,-0.7) {G};
    \node[circle, font=\footnotesize] (13) at (2.25,-0.5) {\$};
    \node[circle, font=\footnotesize] (14) at (1.05,-0.5) {C};
    \node[circle, font=\footnotesize] (15) at (0.7,-1.5) {A};
    \node[circle, font=\footnotesize] (16) at (0.4,-1.8) {G};
    \node[circle, font=\footnotesize] (17) at (1.7,-1.5) {C};
    \node[circle, font=\footnotesize] (18) at (-1.85,-2.35) {A};
    \node[circle, font=\footnotesize] (19) at (-2,-2.75) {G};
    \node[circle, font=\footnotesize] (20) at (-2,-3.15) {C};
    \node[circle, font=\footnotesize] (21) at (-1.9,-3.55) {G};
    \node[circle, font=\footnotesize] (22) at (-1.55,-4.4) {A};
    \node[circle, font=\footnotesize] (23) at (-1.55,-4.8) {G};
    \node[circle, font=\footnotesize] (24) at (-1.55,-5.2) {A};
    \node[circle, font=\footnotesize] (25) at (-1.55,-5.6) {G};
    \node[circle, font=\footnotesize] (26) at (-1.55,-6) {C};
    \node[circle, font=\footnotesize] (27) at (-1.55,-6.4) {G};
    \node[circle, font=\footnotesize] (28) at (-1.55,-6.8) {C};
    \node[circle, font=\footnotesize] (29) at (-1.45,-7.2) {G};
    \node[circle, font=\footnotesize] (30) at (-1.2,-7.55) {C};
    \node[circle, font=\footnotesize] (31) at (-0.85,-7.85) {\$};
    \node[circle, font=\footnotesize] (32) at (-0.3,-3.2) {C};
    \node[circle, font=\footnotesize] (33) at (-0.55,-3.5) {G};
    \node[circle, font=\footnotesize] (34) at (-1,-2.35) {C};
    \node[circle, font=\footnotesize] (35) at (-1,-2.75) {G};
    \node[circle, font=\footnotesize] (36) at (0.3,-3) {A};
    \node[circle, font=\footnotesize] (37) at (0.3,-3.4) {G};
    \node[circle, font=\footnotesize] (38) at (0.3,-3.8) {C};
    \node[circle, font=\footnotesize] (39) at (0.3,-4.2) {G};
    \node[circle, font=\footnotesize] (40) at (0.3,-4.6) {C};
    \node[circle, font=\footnotesize] (41) at (0.3,-5.0) {G};
    \node[circle, font=\footnotesize] (42) at (0.3,-5.4) {C};
    \node[circle, font=\footnotesize] (43) at (0.3,-5.8) {\$};
    \node[circle, font=\footnotesize] (44) at (-0.9,-4.4) {C};
    \node[circle, font=\footnotesize] (45) at (-0.75,-4.75) {G};
    \node[circle, font=\footnotesize] (46) at (-0.65,-5.1) {C};
    \node[circle, font=\footnotesize] (47) at (-0.55,-5.45) {\$};
    \node[circle, font=\footnotesize] (48) at (1.85,-2.8) {\$};
    \node[circle, font=\footnotesize] (49) at (1.9,-4.6) {A};
    \node[circle, font=\footnotesize] (50) at (1.75,-4.85) {G};
    \node[circle, font=\footnotesize] (51) at (1.6,-5.05) {A};
    \node[circle, font=\footnotesize] (52) at (1.45,-5.3) {G};
    \node[circle, font=\footnotesize] (53) at (1.3,-5.525) {C};
    \node[circle, font=\footnotesize] (54) at (1.15,-5.75) {G};
    \node[circle, font=\footnotesize] (55) at (1,-5.975) {C};
    \node[circle, font=\footnotesize] (56) at (0.825,-6.2) {G};
    \node[circle, font=\footnotesize] (57) at (0.675,-6.45) {C};
    \node[circle, font=\footnotesize] (58) at (0.5275,-6.675) {\$};
    \node[circle, font=\footnotesize] (59) at (2.45,-2.95) {G};
    \node[circle, font=\footnotesize] (60) at (2.45,-4.95) {C};
    \node[circle, font=\footnotesize] (61) at (2.4,-6.7) {\$};
    \node[circle, font=\footnotesize] (62) at (1.8,-6.2) {G};
    \node[circle, font=\footnotesize] (63) at (1.55,-6.4) {C};
    \node[circle, font=\footnotesize] (64) at (1.3,-6.6) {\$};
   
    \draw[->] (a) edge (b);
    \draw[->] (a) edge (f);
    \draw[->] (b) edge (c);
    \draw[->] (b) edge (d);
    \draw[->] (c) edge (h);
    \draw[->] (c) edge (i);
    \draw[->] (d) edge (e);
    \draw[->] (e) edge (g);
    \draw[->] (g) edge (i);
    \draw[->] (d) edge (i);
    \draw[->] (e) edge (i);
    \draw[->] (f) edge (h);
    \draw[->] (h) edge (i);
    \draw[->] (f) to[out=-135, in=135] (h);
    \draw[->] (a) to[out=0, in=90] (d);
    \draw[->] (g) to[out=-90, in=0] (i);
    \draw[->] (h) to[out=-90, in=180] (i);
    \draw (a) to[out=0, in=90] (j);
    \draw (j) to[out=90, in=-90] (k);
    \draw[->] (k) to[out=-90, in=0] (i);
    
\end{tikzpicture}

        \end{adjustbox}
        \caption{CDAWG with edge labels.}
    \end{subfigure}
    \begin{subfigure}[t]{0.32\textwidth}
        \centering
        \begin{adjustbox}{max width=1.0\textwidth}

\begin{tikzpicture}
    \node[circle, draw] (a) at (0,0) {};
    \node[circle, draw] (b) at (1.25,-1.25) {};
    \node[circle, draw] (c) at (0,-2.5) {};
    \node[circle, draw] (d) at (2.25,-2.5) {};
    \node[circle, draw] (e) at (2.25,-4.5) {};
    \node[circle, draw] (f) at (-1.25,-2) {};
    \node[circle, draw] (g) at (2.25,-6.25) {};
    \node[circle, draw] (h) at (-1.25,-4) {};
    \node[circle, draw] (i) at (0,-7.75) {};
    \node[inner sep=0pt] (j) at (2.75,-2) {}; 
    \node[inner sep=0pt] (k) at (2.75,-6.5) {}; 
    \node[inner sep=0pt] (x) at (3.5,-3) {}; 
    
    \node[circle, font=\footnotesize] (10) at (-0.6,-0.5) {0,1};
    \node[circle, font=\footnotesize] (12) at (0.15,-0.55) {5,5};
    \node[circle, font=\footnotesize] (13) at (2.4,-0.5) {15,15};
    \node[circle, font=\footnotesize] (14) at (1.35,-0.9) {12,12};
    \node[circle, font=\footnotesize] (15) at (0.65,-1.5) {0,1};
    \node[circle, font=\footnotesize] (17) at (1.3,-1.9) {12,12};
    \node[circle, font=\footnotesize] (18) at (-1.85,-2.35) {2,5};
    \node[circle, font=\footnotesize] (22) at (-1.6,-4.4) {6,15};
    \node[circle, font=\footnotesize] (32) at (-0.3,-3.2) {4,5};
    \node[circle, font=\footnotesize] (34) at (-0.95,-2.35) {4,5};
    \node[circle, font=\footnotesize] (36) at (0.35,-2.95) {8,15};
    \node[circle, font=\footnotesize] (44) at (-0.7,-4.4) {12,15};
    \node[circle, font=\footnotesize] (48) at (1.65,-2.85) {15,15};
    \node[circle, font=\footnotesize] (49) at (1.75,-4.7) {6,15};
    \node[circle, font=\footnotesize] (59) at (2.5,-2.95) {5,5};
    \node[circle, font=\footnotesize] (60) at (1.85,-5.8) {12,12};
    \node[circle, font=\footnotesize] (61) at (1.4,-7.25) {15,15};
    \node[circle, font=\footnotesize] (62) at (1.55,-6.4) {13,15};

    \draw[->] (a) edge (b);
    \draw[->] (a) edge (f);
    \draw[->] (b) edge (c);
    \draw[->] (b) edge (d);
    \draw[->] (c) edge (h);
    \draw[->] (c) edge (i);
    \draw[->] (d) edge (e);
    \draw[->] (e) edge (g);
    \draw[->] (g) edge (i);
    \draw[->] (d) edge (i);
    \draw[->] (e) edge (i);
    \draw[->] (f) edge (h);
    \draw[->] (h) edge (i);
    \draw[->] (f) to[out=-135, in=135] (h);
    \draw[->] (a) to[out=0, in=90] (d);
    \draw[->] (g) to[out=-90, in=0] (i);
    \draw[->] (h) to[out=-90, in=180] (i);
    \draw (a) to[out=0, in=90] (j);
    \draw (j) to[out=90, in=-90] (k);
    \draw[->] (k) to[out=-90, in=0] (i);
\end{tikzpicture}
        \end{adjustbox}
        \caption{CDAWG with string positions.}
    \end{subfigure}
    \begin{subfigure}[t]{0.32\textwidth}
        \centering
        \begin{adjustbox}{max width=1\textwidth}

\begin{tikzpicture}

    \node[rectangle, font=\fontsize{10}{12}\selectfont] (aaa) at (0,0) {A};
    \node[rectangle, font=\fontsize{10}{12}\selectfont] (bbb) at (0.3,0) {G};
    \node[rectangle, font=\fontsize{10}{12}\selectfont] (ccc) at (0.6,0) {A};
    \node[rectangle, font=\fontsize{10}{12}\selectfont] (ddd) at (0.9,0) {G};
    \node[rectangle, font=\fontsize{10}{12}\selectfont] (eee) at (1.2,0) {C};
    \node[rectangle, font=\fontsize{10}{12}\selectfont] (fff) at (1.5,0) {G};
    \node[rectangle, font=\fontsize{10}{12}\selectfont] (ggg) at (1.8,0) {A};
    \node[rectangle, font=\fontsize{10}{12}\selectfont] (hhh) at (2.1,0) {G};
    \node[rectangle, font=\fontsize{10}{12}\selectfont] (iii) at (2.4,0) {A};
    \node[rectangle, font=\fontsize{10}{12}\selectfont] (jjj) at (2.7,0) {G};
    \node[rectangle, font=\fontsize{10}{12}\selectfont] (kkk) at (3,0) {C};
    \node[rectangle, font=\fontsize{10}{12}\selectfont] (lll) at (3.3,0) {G};
    \node[rectangle, font=\fontsize{10}{12}\selectfont] (mmm) at (3.6,0) {C};
    \node[rectangle, font=\fontsize{10}{12}\selectfont] (nnn) at (3.9,0) {G};
    \node[rectangle, font=\fontsize{10}{12}\selectfont] (ooo) at (4.2,0) {C};
    \node[rectangle, font=\fontsize{10}{12}\selectfont] (ppp) at (4.5,0) {\$};

    \node[rectangle, font=\tiny] (ab) at (0,-0.35) {0};
    \node[rectangle, font=\tiny] (ac) at (0.3,-0.35) {1};
    \node[rectangle, font=\tiny] (ad) at (0.6,-0.35) {2};
    \node[rectangle, font=\tiny] (ae) at (0.9,-0.35) {3};
    \node[rectangle, font=\tiny] (af) at (1.2,-0.35) {4};
    \node[rectangle, font=\tiny] (ag) at (1.5,-0.35) {5};
    \node[rectangle, font=\tiny] (ah) at (1.8,-0.35) {6};
    \node[rectangle, font=\tiny] (ai) at (2.1,-0.35) {7};
    \node[rectangle, font=\tiny] (aj) at (2.4,-0.35) {8};
    \node[rectangle, font=\tiny] (ak) at (2.7,-0.35) {9};
    \node[rectangle, font=\tiny] (al) at (3,-0.35) {10};
    \node[rectangle, font=\tiny] (am) at (3.3,-0.35) {11};
    \node[rectangle, font=\tiny] (an) at (3.6,-0.35) {12};
    \node[rectangle, font=\tiny] (ao) at (3.9,-0.35) {13};
    \node[rectangle, font=\tiny] (ap) at (4.2,-0.35) {14};
    \node[rectangle, font=\tiny] (aq) at (4.5,-0.35) {15};
    
    \draw[line width=1pt] (-0.1,-0.65) -- (0.1,-0.65) node[midway,below=6pt] {A};
    \draw[line width=1pt] (0.0,-0.65) -- (0.0,-0.85);
    \draw[line width=1pt] (-0.1,-0.65) -- (-0.1,-0.5);
    \draw[line width=1pt] (0.1,-0.65) -- (0.1,-0.5);
    \draw[line width=1pt] (0.2,-0.65) -- (1.3,-0.65) node[midway,below=9pt] {$\alpha$};
    \draw[line width=1pt] (0.75,-0.65) -- (0.75,-0.85);
    \draw[line width=1pt] (0.2,-0.65) -- (0.2,-0.5);
    \draw[line width=1pt] (1.3,-0.65) -- (1.3,-0.5);
    \draw[line width=1pt] (1.4,-0.65) -- (1.9,-0.65) node[midway,below=6pt] {$\beta$};
    \draw[line width=1pt] (1.65,-0.65) -- (1.65,-0.85);
    \draw[line width=1pt] (1.4,-0.65) -- (1.4,-0.5);
    \draw[line width=1pt] (1.9,-0.65) -- (1.9,-0.5);
    \draw[line width=1pt] (2.0,-0.65) -- (3.1,-0.65) node[midway,below=9pt] {$\alpha$};
    \draw[line width=1pt] (2.55,-0.65) -- (2.55,-0.85);
    \draw[line width=1pt] (2.0,-0.65) -- (2.0,-0.5);
    \draw[line width=1pt] (3.1,-0.65) -- (3.1,-0.5);
    \draw[line width=1pt] (3.2,-0.65) -- (3.7,-0.65) node[midway,below=8pt] {$\gamma$};
    \draw[line width=1pt] (3.45,-0.65) -- (3.45,-0.85);
    \draw[line width=1pt] (3.2,-0.65) -- (3.2,-0.5);
    \draw[line width=1pt] (3.7,-0.65) -- (3.7,-0.5);
    \draw[line width=1pt] (3.8,-0.65) -- (4.3,-0.65) node[midway,below=8pt] {$\gamma$};
    \draw[line width=1pt] (4.05,-0.65) -- (4.05,-0.85);
    \draw[line width=1pt] (3.8,-0.65) -- (3.8,-0.5);
    \draw[line width=1pt] (4.3,-0.65) -- (4.3,-0.5);
    \draw[line width=1pt] (4.4,-0.65) -- (4.6,-0.65) node[midway,below=6pt] {\$};
    \draw[line width=1pt] (4.5,-0.65) -- (4.5,-0.85);
    \draw[line width=1pt] (4.4,-0.65) -- (4.4,-0.5);
    \draw[line width=1pt] (4.6,-0.65) -- (4.6,-0.5);

    \node[rectangle, font=\fontsize{12}{12}\selectfont] (1) at (1.4,-4.5) {$S$};
    \node[rectangle, font=\fontsize{12}{12}\selectfont, align=right] (2) at (3,-4.5) {A$\alpha \beta \alpha \gamma \gamma \$$};
    \draw[->] (1) to[out=0, in=180] node[pos=0.3, above, sloped, font=\footnotesize] {} (2);
    \node[rectangle, font=\fontsize{12}{12}\selectfont] (3) at (1.4,-5) {$\alpha$};
    \node[rectangle, font=\fontsize{12}{12}\selectfont, align=right] (4) at (2.35,-5) {$\beta \gamma$};
    \draw[->] (3) to[out=0, in=180] node[pos=0.3, above, sloped, font=\footnotesize] {} (4);
    \node[rectangle, font=\fontsize{12}{12}\selectfont] (5) at (1.4,-5.5) {$\beta$};
    \node[rectangle, font=\fontsize{12}{12}\selectfont, align=right] (6) at (2.4,-5.5) {GA};
    \draw[->] (5) to[out=0, in=180] node[pos=0.3, above, sloped, font=\footnotesize] {} (6);
    \node[rectangle, font=\fontsize{12}{12}\selectfont] (7) at (1.4,-6) {$\gamma$};
    \node[rectangle, font=\fontsize{12}{12}\selectfont, align=right] (8) at (2.4,-6) {GC};
    \draw[->] (7) to[out=0, in=180] node[pos=0.3, above, sloped, font=\footnotesize] {} (8);

    \node[inner sep=0pt] (k) at (2.75,-7.0) {}; 
    \draw[decorate,decoration={brace,amplitude=10pt},line width=1pt] (1.2,-6.25) -- (1.2,-4.25) node[midway,left=12pt, font=\fontsize{12}{12}\selectfont] {$R =$};
    \node[rectangle, font=\fontsize{12}{12}\selectfont] (7) at (1.18,-2.8) {$\Sigma = \{\text{A},\text{C}, \text{G}\}$};
    \node[rectangle, font=\fontsize{12}{12}\selectfont] (7) at (1.31,-3.4) {$V = \{S, \alpha, \beta, \gamma\}$};
    \node[rectangle, font=\fontsize{12}{12}\selectfont] (7) at (1.46,-2.2) {$G_T=\langle V,\Sigma,R,S \rangle$};
\end{tikzpicture}

        \end{adjustbox}
        \caption{A straight-line grammar.}
    \end{subfigure}
\end{center}
\caption{The CDAWG and a straight-line grammar for string $T = \text{AGAGCGAGAGCGCGC}\$$.}\label{fig:example}
\end {figure}

A \emph{context-free grammar} (CFG) is a set of recursive rules that describe how to form strings from a language's alphabet.
Formally, a CFG is defined as $G=\langle V,\Sigma,R,S \rangle$, where $V$ is a finite set of non-terminal characters, $\Sigma$ is a finite set of terminal characters disjoint from $V$, $R$ is a finite relation in $V \times (V \cup \Sigma)^*$, and $S$ is the symbol in $V$ that should be used as the \emph{start rule} when using $G$ to parse or generate a string.
$R$ defines the \emph{rules} of the grammar and each rule is written as $v \rightarrow (V \cup \Sigma)^*$, where $v \in V$.
A rule may be referred to by the non-terminal character $v$ on its left side, and the right side of a rule is called the rule's \emph{production}.
We denote the production of a non-terminal character $v$ as $R[v]$ and we denote a non-terminal that occurs in the start rule's production as $v \in R[S]$.
A \emph{straight-line grammar} (SLG) is a CFG that unambiguously produces exactly one string.
We denote an SLG that produces only string $T$ as $G_T$.
We define $N=\sum_{v \in V}|R[v]|$ as the size of $G_T$ measured as the sum of the lengths of all of the productions in $G_T$.
We use $H$ to denote the maximum depth, or \emph{height}, of $G_T$.
See Figure~\ref{fig:example} for an example SLG.

\subsection{Random Access and Pattern Matching}

\emph{Random access} is the ability to efficiently access arbitrary elements of a collection.
For string $T$, we denote random access to a single character as $T[i]$ and we denote random access to a substring as $T[i..j]$, where $0 \leq i, j < n$ and $i<j$.
Similarly, for an SLG $G_T$ that produces $T$, we denote random access to character $T[i]$ as $G_T[i]$ and we denote random access to a substring $T[i..j]$ as $G_T[i..j]$.
We denote the time-complexity for random access on $G_T$ as $\bigO{\text{ra}(r)}$, where $r$ is the number of characters accessed.

Given a pattern $P$, where $m=|P|$, and a representation/index of a text $T$, \emph{pattern matching} is composed of three related problems:
1) determine if $P$ exists in $T$, 2) count the number of occurrences of $P$ in $T$, and 3) retrieve the positions of the occurrences of $P$ in $T$.
(1) is known as the \emph{decision version} of the pattern matching problem.
In this work, ``pattern matching'' refers to all three problems unless stated otherwise.
We use $\text{occ}$ to denote the number of occurrences of a pattern $P$ in text $T$.

\section{Related work}
\label{sec:related}

Pattern matching on grammar-compressed strings is an active area of research.
However, \cite{claude2021logarithmic} appears to currently be the only approach that can be used on any SLG.
In \cite{claude2021logarithmic} the authors present a novel search index that requires $\bigO{N\lg{n}}$ additional space and enables pattern matching in $\bigO{(m^2 + \text{occ})\lg{N}}$ time.
Conversely, our approach matches patterns in \timeComplexity\ time while requiring only \spaceComplexity\ additional space.
Critically, \cite{claude2021logarithmic} uses their own specific algorithm for random access, whereas our approach is compatible with any random access technique.
This allows for application-specific trade-offs between run-time and space and does not require any additional space when computing on grammars that naturally support random access.

There has been a variety of recent techniques for pattern matching on specific types of SLGs that are worth noting.
In \cite{ganardi2022pattern} the authors present the first linear-time pattern matching algorithm that solves the decision version of the problem in $\bigO{N+m}$ time.
Although they claim their approach can be used on any SLG, the authors rely on the technique of \cite{ganardi2021balancing} to reduce the input SLG to a balanced straight-line program with depth $\bigO{\log{N}}$, which may result in a grammar that is actually larger than the input SLG.
In \cite{akagi2021grammar} the authors present a method for pattern matching on SLGs built using induced suffix sorting~\cite{nunes2018induced} that can match patterns in $\bigO{m\lg{|V \cup \Sigma|}+\text{occ}_C\lg{|V \cup \Sigma|}\lg{n}+occ}$ time, where $\text{occ}_C$ is the number of occurrences of a chosen ``core'' $C$ of $P$ in the right hand side of all production rules of the grammar.
Similarly, in \cite{deng2022fm-indexing} the authors present a method for pattern matching on SLGs built using induced suffix sorting by FM-indexing the grammar.
This approach has the same asymptotic run-time complexity as the FM-index with a wavelet tree encoding the run-length burrows-wheeler transform.
Lastly, in \cite{christiansen2021optimal-time} the authors describe the first self-indexes able to count and locate pattern occurrences in optimal $\bigO{m+\text{occ}}$ time within a space bounded by the size of the most popular dictionary compressors.
Unfortunately, this work is purely theoretical and only applies to SLGs based on locally-consistent parsing~\cite{mehlhorn1997maintaining}.
\cite{kociumaka2023near-optimal} improves upon this result by further reducing the space bound.

In addition to enabling pattern matching on any SLG, a distinguishing feature of our approach is that the CDAWG can be used for a variety of string tasks that have not yet been thoroughly studied in the SLG literature.
However, the CDAWG has technically already been used to enable these tasks on a few very specific SLGs.
In recent work, SLGs have been used as an alternative to explicitly storing $T$ to asymptotically reduce the space required by the CDAWG.
In \cite{takagi2017linear-size,cdawg2017belazzougui,inenaga2024linearsize} the authors achieve an $\bigO{el(T) + er(T)}$ space bound, and in \cite{belazzougui2017fast} the authors achieve an $\bigO{er(T)}$ space bound, which is currently the best know space bound for the CDAWG.
These SLGs are induced from the structure of the CDAWG and tend to achieve poor compression in practice~\cite{cleary2023dcc}.

\section{The CDAWG Index and Grammar-Compressed Strings}
\label{sec:idea}

We propose to use the CDAWG to index the SLGs of grammar-compressed strings.
Specifically, we propose to replace the CDAWG's copy of the indexed string $T$ with an SLG $G_T$ that produces $T$.
Critically, $G_T$ should support random access so that it may be used directly in place of $T$ without requiring modifications to the CDAWG.
There are four factors motivating this approach:
\begin{enumerate}
    \item Random access is a fundamental string operation that we believe any pragmatic grammar compressor should support in conjunction with pattern matching.
    \item Decoupling random access from the pattern matching task allows for application-specific trade-offs between space and time complexity.
    \item Random access to SLGs is an active area of research\cite{gagie2020practical,bille2015random,nunes2022grammar}; by taking a generalized approach we are ensuring that this result will remain compatible with the state of the art.
    \item The CDAWG is invariant relative to the SLG it indexes; the CDAWG can be computed independently of an SLG and be used on different SLGs that produce the same string.
\end{enumerate}
Regardless of the invariant property, the CDAWG can be built directly from the SLG that it indexes.

\section{Algorithms and Data Structures}
\label{sec:algorithms}

In this Section we present the algorithms and data structures that implement our approach.

\subsection{SLG Representation}

Since our approach works with any SLG, we opted to load grammars into the canonical representation of an SLG: an array or arrays.
In this representation each subarray is a rule in the grammar and its values are integers representing the rule's non-terminal characters (array indexes) and terminal characters (integers less than 256 are cast to their respective ASCII characters).
This representation requires $o(N)$ space.

\subsection{Random Access}

Rather than use an existing algorithn for random access, we opted to use a na\"ive algorithm that is easy to understand and implement.
Our approach is to index the position in the original text of each (non-)terminal character in the start rule.
Specifically, the positions are the keys in a sorted associative array and the values are the corresponding (non-)terminal characters in the start rule.
This index requires $\bigO{|R[S]|}$ additional space and takes $\bigO{n}$ time to compute since the entire parse tree must be traversed to compute the start rule character positions.
This run-time can be improved to $\bigO{N}$ time with memoization.

Using this associative array, a random access query is then a matter of performing a binary search to find the largest key that is less-than-or-equal-to the start position of the query and decoding the grammar from there while ignoring the characters that precede the start position of the query.
Let $l_{max}=\max(\{|R[v]| : v \in V \land v \in R[S]\})$.
It follows that the worst case random access run-time is $\bigO{\lg |R[S]| + l_{max} \lg H + s\lg H}$, where $\bigO{\lg |R[S]|}$ is for the binary search, $\bigO{l_{max} \lg H}$ is for ignoring characters, and $\bigO{s\lg H}$ is for decoding $s$ characters after the start position.

We acknowledge that this approach to random access is not the state of the art in terms of asymptotic space or time.
Regardless, our experiments show that even this na\"ive approach can achieve state of the art run-time performance.

\subsection{CDAWG Index Construction}
\label{sec:algs:cdawg}

For CDAWG construction, we use the algorithm of \cite{inenaga2005online}.
It is an on-line algorithm that updates the CDAWG with each character read from the text.
This makes it particularly well suited for SLGs because the SLG can be decoded as the algorithm progresses, rather than requiring the entire text up-front.

One nuance to be aware of is that when adding a new character the algorithm requires random access to characters already present in the CDAWG.
If your random access algorithm has efficient single character performance, then implementing the algorithm is simply a matter of replacing all occurrences of ``$T[i]$'' in the algorithm with ``$G_T[i]$''.
However, if you are using a random access algorithm with insufficient single character access performance, such as the one we previously described, then special care must be taken.

An effective approach to minimize single character random access queries is to memoize the first character of each CDAWG edge and to use a first-in-first-out cache.
In addition to improving search performance, first character random access queries are quite sporadic and often result in cache misses even when using a more sophisticated cache policy, like least recently used.
Thus, memoization is the preferred approach for avoiding these random access queries.
All remaining single character random accesses tend to occur near the character currently being added.
We found that a simple first-in-first-out cache can handle the majority of these queries.

Memoization and caching aside, the the run-time of the CDAWG construction algorithm of \cite{inenaga2005online} on an SLG $G_T$ is $\bigO{\text{ra}(n)}$ and it requires $\bigO{\text{er}(T)}$ space, i.e. the final size of the constructed CDAWG.

\subsection{SLG Pattern Matching}

For pattern matching, we use the canonical CDAWG pattern matching algorithm of \cite{blumer1987inverted}.
As with the CDAWG index construction algorithm, implementing the algorithm of \cite{blumer1987inverted} is simply a matter of replacing all occurrences of ``$T[i..j]$'' in the algorithm with ``$G_T[i..j]$''.
The run-time of this algorithm on SLG $G_T$ is $\bigO{\text{ra}(m)\log{\sigma}+\text{occ}}$.
This can be improved to \timeComplexity\ by memoizing the first character of each edge in the CDAWG.

\section{Results}
\label{sec:results}

In this Section we describe the implementation of our algorithms and data structures and discuss the experiments we used to evaluate our approach.

\subsection{Implementation}

We implemented our algorithms and data structures in C++.
As previously mentioned, our CDAWG construction and pattern matching algorithms are essentially verbatim implementations of the algorithms by \cite{inenaga2005online} and \cite{blumer1987inverted}, respectively.
The source code is available at \url{https://github.com/alancleary/cdawg-index-cfg}.

\subsection{Experiments}
\label{sec:experiments}

For our primary experiments we generated SLGs for a variety of data sets from the \emph{Pizza\&Chili} corpus.\footnote{\url{https://pizzachili.dcc.uchile.cl/}}
Specifically, we used all data sets from the \emph{real} and \emph{artificial} collections.
For each data set, we generated SLGs using Navarro's reference implementation of Re-Pair\footnote{\url{https://users.dcc.uchile.cl/~gnavarro/software/repair.tgz}} and MR-RePair\cite{furuya2019mrrepair}.
Then, using our software, we computed a na\"ive random access index and a CDAWG index for each SLG and ran benchmarks.
See Table~\ref{table:data} for information about the data sets, CDAWGs, and SLGs.

For each SLG, we generated substrings of length 10, 100, 1,000, and 10,000 at pseudo-random positions in the text and timed how long it took to search for the substrings using the SLG's CDAWG index.
We performed this procedure 1,000 times for each substring length and computed the average run-time.
Experiments were run on a System76 Adder WS Laptop with a 13th Gen Intel Core i9-13900HX 32 core 5.4~Ghz processor and 64~GiB of dual-channel DDR5 4800~MHz memory running Pop!\_OS Linux version 22.04.
The results are shown in Table~\ref{table:results}.

We observe that the size of the CDAWG indexes (measured in number of right-extensions) is relatively large compared to the size of the grammars.
However, when considering the size of the data sets and grammars, the CDAWGs are literally orders of magnitude smaller than the $\bigO{N\lg{n}}$ additional space required by \cite{claude2021logarithmic}, even when considering the space required by our na\"ive random access algorithm.
Furthermore, since every grammar is smaller than its corresponding CDAWG, every CDAWG-grammar pair is within the best known $\bigO{\text{er}(T)}$ space asymptotic bound for the CDAWG~\cite{belazzougui2017fast}.

Overall, the CDAWG index had very similar performance on both the Navarro Re-Pair and MR-RePair grammars, despite the MR-RePair grammars being consistently smaller and not as deep.
With the exception of the \emph{cere} and \emph{para} data sets, the CDAWG exhibited substantial run-time performance and scalability of pattern matching.

The distinction between \emph{cere}/\emph{para} and the other data sets is the number of random accesses performed per pattern matching query.
For the length 10,000 patterns, the average number of random accesses was 7,707 on \emph{cere} and 1,223 on \emph{para}.
Conversely, the next largest number of random accesses was 23 on \emph{Escherichia\_Coli}.
The \emph{cere} and \emph{para} data sets are composed of 37 sequences of \emph{Saccharomyces cerevisiae} and 36 sequences of \emph{Saccharomyces paradoxus}, respectively.
While DNA has high information content and is notoriously difficult to compress\cite{kryukov2020scb}, the \emph{Escherichia\_Coli} data set is composed of 23 sequences of \emph{Escherichia coli} and the \emph{influenza} data set is composed of 8,041 sequences of \emph{Haemophilus influenzae}, suggesting that DNA is not the issue, per se, but rather the \emph{cere} and \emph{para} sequences are of poor quality.

To test this hypothesis, we performed the same experiments on the \emph{Yeast Population Reference Panel} (YPRP)\cite{yue2017yprp}.
This corpus contains 7 sequences of \emph{Saccharomyces cerevisiae} and 5 sequences of \emph{Saccharomyces paradoxus}.
These sequences are complete \emph{de novo} genome assemblies of extremely high quality and should contain very few errors from sequencing and assembly.
See Table~\ref{table:data} for information about these data sets and see Table~\ref{table:results} for the experiment results.

We found that the CDAWG indexes for the YPRP \emph{Saccharomyces cerevisiae} and \emph{Saccharomyces paradoxus} data sets were similar in size to the CDAWG indexes for the Pizza\&Chili \emph{cere} and \emph{para} data sets, respectively.
However, the YPRP run-times were similar to those of the other data sets, i.e. not \emph{cere} and \emph{para}, and the average number of random accesses for the length 10,000 patterns were 16 for \emph{Saccharomyces cerevisiae} and 27 for \emph{Saccharomyces paradoxus}.

To confirm that this was due to sequence quality and not the number of sequences, we merged the two YPRP data sets into a \emph{combined} data set.
Although the \emph{combined} data set is still smaller than either the \emph{cere} data set or the \emph{para} data set, the taxonomic distance between \emph{Saccharomyces cerevisiae} and \emph{Saccharomyces paradoxus} should be greater than the distance between any two genomes in either data set.
This is reflected in the fact that the CDAWG index for the \emph{combined} data set is larger than both the CDAWG index for the \emph{cere} data set and the CDAWG index for the \emph{para} data set even though the size of the \emph{combined} data set itself is significantly smaller.
As with the YPRP \emph{Saccharomyces cerevisiae} and \emph{Saccharomyces paradoxus} data sets, we found that the run-times for the \emph{combined} data set were similar to those of the other data sets, i.e. not \emph{cere} and \emph{para}, and the average number of random accesses for the legnth 10,000 patterns was 31.

\begin{table}
    \caption{Data sets used from the \emph{Pizze\&Chili} corpus (\emph{real} and \emph{artificial}) and the \emph{Yeast Population Reference Panel} (\emph{YPRP}). \textbf{Data Set} is the names of the data sets and what collection they belong to, \textbf{Size} is the number of characters in each data set, \textbf{CDAWG} is the number of right-extensions in the CDAWGs for the data sets, and \textbf{Navarro Re-Pair} and \textbf{MR-RePair} are information about the grammars built for these data sets. For the grammars, \textbf{Rules} is the number of rules in the grammars, \textbf{Depth} is the maximum depth of the grammars, and \text{Size} is total lengths of the right-hand sides of the rules in each grammar. Note that empty cells mean the algorithm was unable to build a grammar for the data set.}
    \label{table:data}
    \par\vspace{\baselineskip}
    \hspace{-1.5cm}
    \begin{tabular}{|c|l|c|c|c|c|c|c|c|c|}
        \hline
        \multicolumn{2}{|c|}{\multirow{2}*{\textbf{Data Set}}} & \multirow{2}*{\textbf{Size}} & \multirow{2}*{\textbf{CDAWG}} & \multicolumn{3}{c|}{\textbf{Navarro Re-Pair}} & \multicolumn{3}{c|}{\textbf{MR-RePair}} \\ 
        \cline{5-10}
        \multicolumn{2}{|c|}{} & & & \textbf{Rules} & \textbf{Depth} & \textbf{Size} & \textbf{Rules} & \textbf{Depth} & \textbf{Size} \\
        \hline
        \multirow{9}*{\rotatebox[origin=c]{90}{\emph{real}}} & \emph{Escherichia\_Coli} & 112,689,515 & 31,340,395 & 2,012,087 & 3,279 & 5,625,656 & & & \\
         & \emph{cere} & 461,286,644 & 25,753,919 & 2,561,292 & 1,359 & 5,777,882 & & & \\
         & \emph{coreutils} & 205,281,778 & 9,903,928 & 1,821,734 & 43,728 & 3,796,814 & 437,054 & 30 & 2,423,962 \\
         & \emph{einstein.de.txt} & 92,758,441 & 238,338 & 49,949 & 269 & 112,563 & 21,787 & 42 & 84,392 \\
         & \emph{einstein.en.txt} & 467,626,544 & 951,878 & 100,611 & 1,355 & 263,695 & 49,565 & 48 & 212,824 \\
         & \emph{influenza} & 154,808,555 & 17,163,707 & 643,587 & 366 & 2,174,010 & 429,027 & 28 & 1,986,529 \\
         & \emph{kernel} & 257,961,616 & 5,929,488 & 1,057,914 & 5,822 & 2,185,255 & 246,596 & 34 & 1,373,880 \\
         & \emph{para} & 429,265,758 & 33,847,225 & 3,093,873 & 487 & 7,335,396 & & & \\
         & \emph{world\_leaders} & 46,968,181 & 1,590,565 & 206,508 & 463 & 507,343 & 100,293 & 30 & 407,619 \\
        \hline
        \multirow{3}*{\rotatebox[origin=c]{90}{\emph{artificial}}} & \emph{fib41} & 267,914,296 & 80 & 38 & 40 & 79 & 38 & 40 & 79 \\
         & \emph{rs.13} & 216,747,218 & 262 & 66 & 47 & 156 & 55 & 45 & 147 \\
         & \emph{tm29} & 268,435,456 & 212 & 75 & 45 & 156 & 51 & 29 & 132 \\
        \hline
        \multirow{3}*{\rotatebox[origin=c]{90}{\emph{YPRP}}} & \emph{Saccharomyces cerevisiae} & 83,530,741 & 23,633,473 & 1,940,308 & 2,026 & 4,618,271 &  &  &  \\
         & \emph{Saccharomyces paradoxus} & 59,638,719 & 33,090,778 & 2,152,160 & 488 & 5,977,553 &  &  &  \\
         & \emph{combined} & 143,169,461 & 53,422,575 & 3,755,345 & 2,435 & 9,905,715 &  &  &  \\
        \hline
    \end{tabular}
\end{table}

\begin{table}[b]
    \centering
    \caption{Pattern matching run-times for grammars built on data sets from the \emph{Pizze\&Chili} corpus (\emph{real} and \emph{artificial}) and the \emph{Yeast Population Reference Panel} (\emph{YPRP}). \textbf{Data Set} is the names of the data sets and what collection they belong to and \textbf{Navarro Re-Pair} and \textbf{MR-RePair} are run-times for grammars built using these algorithms. For the run-times, \textbf{10}, \textbf{100}, \textbf{1,000}, and \textbf{10,000} are the lengths of the patterns matched. All run-time are in microseconds. Note that empty cells mean the algorithm was unable to build a grammar for the data set.}
    \label{table:results}
    \par\vspace{\baselineskip}
    \begin{tabular}{|c|l|c|c|c|c|c|c|c|c|}
        \hline
        \multicolumn{2}{|c|}{\multirow{2}*{\textbf{Data Set}}} & \multicolumn{4}{c|}{\textbf{Navarro Re-Pair}} & \multicolumn{4}{c|}{\textbf{MR-RePair}} \\ 
        \cline{3-10}
        \multicolumn{2}{|c|}{} & \textbf{10} & \textbf{100} & \textbf{1,000} & \textbf{10,000} & \textbf{10} & \textbf{100} & \textbf{1,000} & \textbf{10,000} \\
        \hline
        \multirow{9}*{\rotatebox[origin=c]{90}{\emph{real}}} & \emph{Escherichia\_Coli} & 4 & 9 & 34 & 218 & & & & \\
         & \emph{cere} & 4 & 74 & 3,884 & 117,657 & & & & \\
         & \emph{coreutils} & 2 & 3 & 19 & 204 & 1 & 3 & 17 & 189 \\
         & \emph{einstein.de.txt} & 2 & 7 & 23 & 180 & 2 & 5 & 21 & 166 \\
         & \emph{einstein.en.txt} & 1 & 3 & 26 & 220 & 1 & 3 & 24 & 211 \\
         & \emph{influenza} & 4 & 6 & 21 & 190 & 4 & 6 & 22 & 192 \\
         & \emph{kernel} & 2 & 4 & 20 & 202 & 2 & 4 & 19 & 189 \\
         & \emph{para} & 4 & 76 & 3,924 & 5,848 & & & & \\
         & \emph{world\_leaders} & 1 & 3 & 19 & 171 & 1 & 3 & 17 & 160 \\
        \hline
        \multirow{3}*{\rotatebox[origin=c]{90}{\emph{artificial}}} & \emph{fib41} & 0 & 4 & 29 & 225 & 0 & 2 & 19 & 146 \\
         & \emph{rs.13} & 0 & 3 & 22 & 184 & 0 & 2 & 20 & 179 \\
         & \emph{tm29} & 1 & 3 & 19 & 190 & 1 & 3 & 18 & 215 \\
        \hline
        \multirow{3}*{\rotatebox[origin=c]{90}{\emph{YPRP}}} & \emph{Saccharomyces cerevisiae} & 4 & 7 & 22 & 194 &  &  &  &  \\
         & \emph{Saccharomyces paradoxus} & 5 & 16 & 32 & 203 &  &  &  &  \\
         & \emph{combined} & 5 & 18 & 34 & 206 &  &  &  &  \\
        \hline
    \end{tabular}
\end{table}

\subsection{Discussion}

In general, our approach exhibits robust performance across different data sets and grammars.
As for the \emph{cere} and \emph{para} data sets, the poor results could potentially be remedied with a less na\"ive random access algorithm that has better single-character run-time performance.
However, the results on the these data sets also highlight potential points of improvement to the CDAWG index for DNA data sets.

The primary issue with the \emph{cere} and \emph{para} data sets is that they contain large runs of the ``N'' character to denote parts of the genomes that could not be resolved when the genomes were assembled.
Every unique substring (i.e. length) of these runs of ``N'' is a maximal repeat.
Due to the one-to-one correspondence between the maximal repeats of a string and the internal nodes of the CDAWG that recognizes the suffixes of the string, this resulted in an additional node being added to the \emph{cere} and \emph{para} CDAWG indexes for each unique substring of the runs of ``N'', and each such node had an out-edge labeled ``N'' leading to the next such node.
This proliferation of ``N'' nodes was further exacerbated by the fact that all of the sequences in the \emph{cere} and \emph{para} data set files were concatenated into a single line.
Runs of ``N'' tend to occur at the beginning and end of sequences, so the concatenation likely created artificial runs that are longer than any run in the original sequences.
Runs of the same character is a problem in general for CDAWGs.
However, in the case of DNA, the problem character ``N'' is known \emph{a priori} and could be better handled via a special run-length node.
Matching a run of ``N'' characters would then require no random access queries.

Another DNA-specific problem that is prevalent even in high-quality genome assemblies is sheer probability.
The DNA alphabet is so small that for a sufficiently long sequence it is likely that all possible substrings over the alphabet up to a certain length occur in the sequence.
Similar to the runs of ``N'', this results in many nodes connected by edges whose labels are a single character.
Unlike the runs of ``N'', which form chains in the CDAWG, these nodes occur near the source node and are fully connected up to a certain depth.
For sufficiently long patterns, there will likely be no quantifiable affect on pattern matching run-time performance.
However, for shorter patterns the affect is quantifiable even on smaller genomes.
This is likely the cause of the slightly slower run-times for the smaller patterns on the DNA data sets in our experiments.
For longer genomes and many short patterns, the compounded affect could be significant.
This scenario is commonplace in bioinformatics where millions of short-read DNA sequences are aligned to a \emph{Homo sapien} genome in a single analysis.
Perhaps this could be addressed by replacing this portion of the CDAWG with a simple lookup table.
Although this would not reduce the size of the CDAWG asymptotically, matching pattern preffixes using the lookup table would require no random access queries.

\section{Conclusion}
\label{sec:conclusion}

In this work, we proposed using the CDAWG to index the SLGs of grammar-compressed strings.
We showed that when the SLGs being indexed support random access, this approach is quite simple conceptually and easy to implement.
While this approach enables all analyses supported by the CDAWG on any grammar-compressed string, we specifically considered pattern matching.
Our experiments show that even when using a na\"ive random access algorithm, the CDAWG index achieves state of the art run-time performance for pattern matching on grammar-compressed strings.
Additionally, we found that all of the grammars computed for our experiments are smaller than the number of right-extensions in the string they produce and, thus, their CDAWGs are within the best known $\bigO{\text{er}(T)}$ space asymptotic bound.
It would be interesting if this space bound could be generalized to a certain class of SLG, if not all of them.

\begin{credits}
\subsubsection{\ackname} 
The work of Alan M. Cleary, Joseph Winjum, and Jordan Dood was support by NSF award number 2105391.
The work of Shunsuke Inenaga was supported by JSPS KAKENHI grant numbers JP	20H05964, JP23K24808, JP23K18466.

\end{credits}

\clearpage

\bibliographystyle{splncs04}
\bibliography{main}

\end{document}